# Cloak/anti-cloak interactions

Giuseppe Castaldi,[1] Ilaria Gallina,[1,2] Vincenzo Galdi,[1,*] Andrea Alù,[3,4]
and Nader Engheta[3]

[1]*Waves Group, Department of Engineering, University of Sannio, Corso Garibaldi 107, I-82100, Benevento, Italy*
[2]*Department of Environmental Engineering and Physics, University of Basilicata, Via dell'Ateneo Lucano 10, I-85100 Potenza, Italy*
[3]*Department of Electrical and Systems Engineering, University of Pennsylvania, 200 South 33rd Street, Philadelphia, PA 19104, USA*
[4] *Department of Electrical and Computer Engineering, University of Texas at Austin, Austin, TX 78712, USA*
[*]*Corresponding author: vgaldi@unisannio.it*

**Abstract:** Coordinate-transformation cloaking is based on the design of a metamaterial shell made of an anisotropic, spatially inhomogeneous "transformation medium" that allows rerouting the impinging wave around a given region of space. In its original version, it is generally believed that, in the ideal limit, the radiation *cannot* penetrate the cloaking shell (from outside to inside, and viceversa). However, it was recently shown by Chen *et al*. that electromagnetic fields may actually penetrate the cloaked region, provided that this region contains *double-negative* transformation media which, via proper design, may be in principle used to (partially or totally) "undo" the cloaking transformation, thereby acting as an "anti-cloak." In this paper, we further elaborate this concept, by considering a more general scenario of cloak/anti-cloak interactions. Our full-wave analytical study provides new insightful results and explores the effects of departure from ideality, suggesting also some novel scenarios for potential applications.

## 1. Introduction and background

Invisibility "cloaking" of a (penetrable or impenetrable) object, i.e., its concealment to the electromagnetic (EM) illumination via a shell made of a properly engineered material capable of suppressing the overall (near- and far-field) scattering response, has recently emerged as one of the most intriguing and transformative applications in EM and optical engineering, thanks to the rapid advances in the engineering of special materials and "metamaterials." Among the various approaches, it is worth mentioning the plasmonic cloaking method, based on scattering cancellation, proposed by Alù and Engheta [1], the cloaking based on anomalous localized resonances proposed by Milton, Nicorovici, McPhedran and co-workers [2], and the metamaterial cloaking based on coordinate transformations proposed by Pendry, Schurig and Smith [3,4] and by Leonhardt and Philbin [5,6] (experimentally demonstrated at microwave frequencies [7]). The reader is referred to [8] (and references therein) for a recent comparative review of these various approaches.

The coordinate-transformation approach relies on the intuitive geometric idea of designing a material shell capable of suitably *bending* the ray trajectories (which describe the high-frequency power flux) around the object to be cloaked. The design may be performed by first deriving the desired field distribution in a fictitious *curved-coordinate* space containing a "hole," and subsequently exploiting the formal invariance of Maxwell's equations under coordinate transformations to translate such distribution into a conventionally flat, Cartesian space, filled by a suitably anisotropic and spatially inhomogeneous transformation medium [3-6].

Interestingly, the range of applicability of this approach is not restricted to the asymptotic ray-optical limit. In fact, *exact* full-wave analytic studies in the spherical [9] and cylindrical [10] canonical geometries, based on suitable generalizations of the Mie series, have demonstrated the possibility of achieving, in principle, *perfect cloaking*, i.e., zero external scattering and zero transmission into the cloaked region, at *any* given frequency. Thus, for an "ideal" cloaking (implying a lossless, non-dispersive, anisotropic, spatially inhomogeneous metamaterial, with extreme values of the relative permittivities and permeabilities ranging from zero to infinity) of an isotropic object, the field *cannot* penetrate from outside to inside, and vice-versa [11]. The reader is referred to, e.g., [12-15] for the effects of perturbations and simplifications/reductions in the metamaterial parameters, and to [16] for the bandwidth limitations.

In a recent paper [17], Chen *et al.* introduced a new twist in this concept, showing that the above picture may not be longer valid if the cloaked region is allowed to contain another anisotropic, spatially inhomogeneous medium suitably designed to act as an "anti-cloak." In particular, with reference to a two-dimensional (2-D) cylindrical scenario, they showed that it is possible, in principle, to design a transformation-medium shell (based on a linearly-decreasing radial coordinate transformation) that, when laid directly between the cloak and the object, is capable of "undoing" the cloaking transformation, restoring (partially or totally) the original scattering response. In terms of practical feasibility, the same limitations mentioned above for the cloak clearly hold for the anti-cloak as well, with a further complication given by the *double negative* (DNG) [18] character of the anti-cloak transformation-medium, dictated by the monotonic negative slope of the corresponding coordinate-transformation [17].

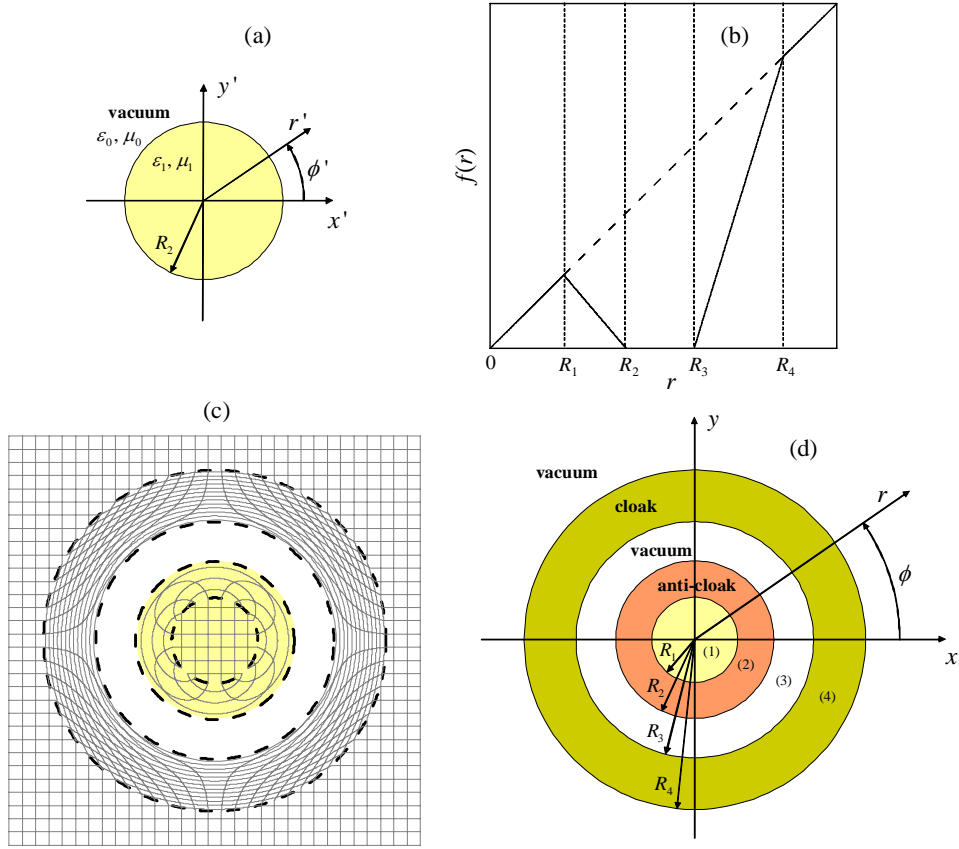

Fig. 1. Geometry of the problem. (a) Homogeneous circular cylinder in the auxiliary space. (b) Radial coordinate transformation, as in (2). (c) Topological interpretation of the mapping, with curved coordinates in the transformed regions. (d) Alternative interpretation of the mapping in a globally flat, Cartesian space, with cloak- and anti-cloak-type transformation media.

In this paper, we elaborate upon the interactions between a cloak and an anti-cloak, by considering a more general scenario where: i) the inner cylinder is made of an arbitrary isotropic, homogeneous metamaterial, and ii) the cloak and the anti-cloak are not directly contiguous, but they may be separated by a vacuum shell. Via a full-wave analytical study, we show that, depending on the nature of the inner object, the anti-cloak-type effect may also be achieved with *double-positive* (DPS), or *single-negative* (SNG) transformation media, relaxing the complexity of the anti-cloak design and making its practical realization arguably more feasible. Moreover, we show that the effect may be preserved even in the presence of a vacuum shell separating cloak and anti-cloak, suggesting a counterintuitive "field-tunneling" mechanism. Finally, we explore the effects of the presence of penetrable objects inside the vacuum shell, as well as of the departure from ideal conditions, i.e., the presence of losses, and deviations from the ideal design parameters. These results provide new insights in the anti-cloaking mechanism, and may ensure further degrees of freedom in the design of an anti-cloak. Moreover, besides the potential application (as a cloaking cancellation) suggested in

[17], we envision other potentially interesting configurations where cloaked and field-accessible regions may coexist.

The rest of the paper is organized as follows: in Sect. 2 we illustrate the problem geometry and formulation; in Sect. 3, we outline the full-wave analytical solution; in Sect. 4, we illustrate some representative results. Finally, in Sect. 5, we provide some concluding remarks and hints for future research.

**2. Problem geometry and formulation**

Referring to Fig. 1(a), we start considering a 2-D scenario consisting of an isotropic, homogeneous circular cylinder of radius $R_2$, with dielectric permittivity $\varepsilon_1$ and magnetic permeability $\mu_1$, immersed in vacuum in the auxiliary space $(x',y',z')$. In the associated cylindrical $(r',\phi',z')$ reference system, this configuration may be parameterized by the permittivity and permeability distributions

$$\varepsilon'(r') = \begin{cases} \varepsilon_1, & 0 < r' < R_2 \\ \varepsilon_0, & r' > R_2 \end{cases}, \quad \mu'(r') = \begin{cases} \mu_1, & 0 < r' < R_2 \\ \mu_0, & r' > R_2 \end{cases}, \tag{1}$$

with $\varepsilon_0$ and $\mu_0$ denoting the vacuum dielectric permittivity and magnetic permeability, respectively. We map the above configuration onto the actual physical space $(x,y,z)$ via the piecewise linear radial (in the associated $(r,\phi,z)$ cylindrical reference system) coordinate transformation (see Fig. 1(b)):

$$r' = f(r) = \begin{cases} r, & r < R_1, \ r > R_4, \\ R_1\left(\dfrac{R_2 + \Delta_2 - r}{R_2 + \Delta_2 - R_1}\right), & R_1 < r < R_2, \\ R_4\left(\dfrac{r - R_3 + \Delta_3}{R_4 - R_3 + \Delta_3}\right), & R_2 < r < R_4. \end{cases} \tag{2}$$

In the region $R_2 < r < R_3$, an arbitrary coordinate mapping $f(r) < 0$ can be assumed, in light of the theory developed in what follows, and therefore it has not been specified in (2) and in Fig. 1(b). When the (negligibly small) parameters $\Delta_2$ and $\Delta_3$ are zero, it is readily recognized that the outermost layer $R_3 < r < R_4$ corresponds to a standard invisibility cloak [3], whereas the internal layer $R_1 < r < R_2$ corresponds to the anti-cloak introduced in [17]. As is well known, the cloak transformation compresses the cylindrical region $r < R_4$ into the concentric annular layer $R_3 < r < R_4$, thereby creating a "hole" of radius $R_3$ wherein an object may be concealed. The anti-cloak transformation may be interpreted as a somehow *reverse* operation, where an "anti-hole" of radius $R_2 > R_1$ is created around a cylinder of radius $R_1$. From a topological viewpoint, their combination results in a four-layer cylindrical configuration of radii $R_\nu, \nu = 1,...,4$ (see Fig. 1(c)), where the transformed regions $R_3 < r < R_4$ (cloak) and $R_1 < r < R_2$ (anti-cloak) are characterized by *curved* coordinates, while the regions $r < R_1$ and $r > R_4$ maintain the flat, Cartesian metrics. Conversely, the layer $R_2 < r < R_3$ does not admit any physical image (i.e., $r' > 0$) in the auxiliary $(x',y',z')$ space, thereby constituting a "cloaked" region *inaccessible* to the EM fields. It is instructive to observe in Fig. 1(c) how the curved coordinates in the cloak and anti-cloak regions bend in a somehow complementary fashion around the cloaked layer, while matching the flat, Cartesian coordinates in the regions $r < R_1$ and $r > R_4$.

Within the framework of transformation optics, invoking the formal invariance of Maxwell's equations under coordinate transformations, the above behavior can be equivalently obtained in a *globally flat* space by filling up the transformed regions with anisotropic, spatially inhomogeneous transformation media, whose permittivity and permeability tensors may be readily derived from the Jacobian matrix of the transformation (2) (see [4] for details). Restricting our attention to transversely-magnetic (TM) polarized fields (i.e., magnetic field parallel to the coaxial cylinders), the relevant components (in cylindrical coordinates) may be expressed as [15]:

$$\varepsilon_r(r) = \varepsilon'(r')\left(\frac{r'}{r}\right)\left[\frac{df(r)}{dr}\right]^{-1}, \quad \varepsilon_\phi(r) = \varepsilon'(r')\left(\frac{r}{r'}\right)\frac{df(r)}{dr}, \quad \mu_z(r) = \mu'(r')\left(\frac{r'}{r}\right)\frac{df(r)}{dr}. \quad (3)$$

Accordingly, in the assumed scenario (see Fig. 1(d)), the layers $R_3 < r < R_4$ (cloak) and $R_1 < r < R_2$ (anti-cloak) are filled by the transformation media arising from (3), while the regions $r < R_1$ and $r > R_4$ remain filled by isotropic, homogeneous media (with constitutive parameters $\varepsilon_1$, $\mu_1$ and $\varepsilon_0$, $\mu_0$, respectively), and the layer $R_2 < r < R_3$ is filled by vacuum. A few considerations are in order here:

i) Note that, as in [10,12], we introduced in the transformation (2) two small dimensional parameters $\Delta_2$ and $\Delta_3$, which parameterize the departure of the cloak and anti-cloak from the ideal case ($\Delta_2, \Delta_3 \to 0$). These are necessary in the design of the anti-cloak, as it will become clear in the following.

ii) In our configuration, unlike that in [17], the cloak and anti-cloak are separated by a vacuum layer $(R_2 < r < R_3)$, which, as observed above, constitutes the cloaked region in the limit $\Delta_2, \Delta_3 \to 0$. In this same limit, the external cloak is perfectly matched with vacuum at the interface $r = R_4$, and the anti-cloak is perfectly matched with the inner medium at the interface $r = R_1$; this yields *zero scattering* outside the region $r > R_4$. Moreover, as a consequence of the *non-monotonic* behavior of the transformation (2), the interfaces bounding the cloaked layer $r = R_2$ and $r = R_3$ are imaged in the same point $r' = 0$ in the auxiliary space (see Fig. 1(b)). This suggests the possibility of an intriguing and somehow counterintuitive mechanism of field transfer between these interfaces, via the "tunneling" through the cloaked layer.

iii) In view of the negative slope of the transformation (2) in the anti-cloak layer $R_1 < r < R_2$, the corresponding constitutive parameters (3) are *opposite in sign* to those of the inner cylinder $(\varepsilon_1, \mu_1)$. This suggests *four* possible configurations of interest, involving the possible combinations of DPS and DNG, or alternatively epsilon-negative (ENG) and mu-negative (MNG), media.

**3. Full-wave analytical solution**

We now consider the scattering of a time-harmonic $(\exp(-i\omega t))$, TM-polarized plane wave, with unit-amplitude $z$-directed magnetic field, impinging from the positive $x$-direction on the four-layer cloak/anti-cloak configuration in Fig. 1(d). It is expedient to represent the incident magnetic field in the associated $(r,\phi,z)$ cylindrical coordinate system in terms of a Fourier-Bessel series:

$$H_z^{(inc)}(r,\phi) = \exp(ik_0 x) = \sum_{n=-\infty}^{\infty} i^n J_n(k_0 r)\exp(in\phi), \quad (4)$$

where $k_0 = \omega\sqrt{\varepsilon_0 \mu_0} = 2\pi/\lambda_0$ denotes the vacuum wavenumber (with $\lambda_0$ being the corresponding wavelength), and $J_n$ denotes the $n$th-order Bessel function of the first kind [19, Chap. 9]. Following [10], we can expand the fields in the various regions by mapping (via (2)) the simple Fourier-Bessel series solution into the auxiliary $(x',y',z')$ space (plane-wave scattering by a homogeneous, isotropic circular cylinder, cf. Fig. 1(a)). Introducing, for notational convenience the "dummy" parameters $R_0 = 0$ and $R_5=\infty$, the Fourier-Bessel expansion of the above field can be compactly written as:

$$H_z(r,\phi) = \sum_{n=-\infty}^{\infty} \left\{ \left(a_n^{(\nu)} + \delta_{\nu 5} i^n\right) J_n[g(r)] + b_n^{(\nu)} Y_n[g(r)] \right\} \exp(in\phi),$$
$$R_{\nu-1} < r < R_\nu, \quad \nu = 1,..,5, \quad (5)$$

where $g(r) = k_0 r$ in the cloaked layer $R_2 < r < R_3$, and $g(r) = \omega\sqrt{\varepsilon'[f(r)]\mu'[f(r)]}f(r)$ elsewhere. Moreover, $a_n^{(\nu)}$ and $b_n^{(\nu)}$ denote the unknown expansion coefficients (to be computed by enforcing the boundary and tangential-field-continuity conditions), $Y_n$ denotes an $n$th-order Bessel function of the second kind [19, Chap. 9], and $\delta_{pq}$ is the Kronecker delta (accounting for the presence of the incident field (4) in the vacuum region $r > R_4$). Note that the field-finiteness condition at $r = 0$ and the radiation-at-infinity condition imply that $b_n^{(1)} = 0$ and $b_n^{(5)} = ia_n^{(5)}$, respectively. The corresponding electric field components can be readily derived from (5) via the relevant Maxwell's curl equation:

$$E_\phi(r,\phi) = \frac{1}{i\omega\varepsilon_\phi(r)} \frac{\partial H_z(r,\phi)}{\partial r}, \quad E_r(r,\phi) = -\frac{1}{i\omega\varepsilon_r(r)} \frac{\partial H_z(r,\phi)}{r\partial \phi}. \quad (6)$$

By enforcing the continuity of the tangential electric and magnetic fields at the interfaces $r = R_1$ and $r = R_4$, we readily derive

$$a_n^{(5)} = -ib_n^{(4)}, \quad a_n^{(4)} = i^n - ib_n^{(4)}, \quad b_n^{(2)} = 0, \quad a_n^{(1)} = a_n^{(2)}. \quad (7)$$

The continuity conditions at the remaining interfaces $r = R_2$ and $r = R_3$ require particular care since, in the limit of ideal cloak $(\Delta_3 \to 0)$ and anti-cloak $(\Delta_2 \to 0)$, the transformation in (2) vanishes, thereby causing the Bessel functions of second kind in the expansion (5) to exhibit a singular behavior. In this ideal scenario, the anti-cloak design would not be possible, since the inner layer of the cloak is perfectly impenetrable. However, in the limit for which $\Delta_2$ and $\Delta_3$ tend both to zero (but are not exactly zero), it is still possible to tailor the design of a suitable anti-cloak, as discussed in the following.

Analogous to [10], and using (in the limit $\Delta_2, \Delta_3 \to 0$) the small argument approximations of the Bessel functions of first [19, Eq. (9.1.7)] and second kind [19, Eqs. (9.1.8) and (9.1.9)], we obtain for the zeroth-order coefficients:

$$a_0^{(1)} = a_0^{(2)} \sim \frac{2}{\pi k_0^2 R_2 R_3 \Lambda \log\left(\frac{k_0 R_4 \Delta_3}{R_4 - R_3}\right)}, \quad a_0^{(3)} \sim -\frac{Y_1(k_0 R_2)}{k_0 R_3 \Lambda \log\left(\frac{k_0 R_4 \Delta_3}{R_4 - R_3}\right)}, \tag{8}$$

$$b_0^{(3)} \sim \frac{J_1(k_0 R_2)}{k_0 R_3 \Lambda \log\left(\frac{k_0 R_4 \Delta_3}{R_4 - R_3}\right)}, \quad a_0^{(5)} = -i b_0^{(4)} \sim \frac{i\pi}{2\log\left(\frac{k_0 R_4 \Delta_3}{R_4 - R_3}\right)}, \tag{9}$$

and for the $(n \neq 0)$ th-order coefficients:

$$a_n^{(1)} = a_n^{(2)} \sim \frac{4|n|i^n \varepsilon_1}{\pi k_0 R_3 \Omega_n} \left[\frac{R_4 (R_2 - R_1) \Delta_3}{\sqrt{\varepsilon_1 \mu_1} R_1 (R_4 - R_3) \Delta_2}\right]^{|n|}, \tag{10}$$

$$a_n^{(3)} \sim \frac{2^{1-|n|} i^{|n|} \Psi_n^{(2)}(k_0 R_2)}{(|n|-1)! k_0 R_3 \Omega_n} \left(\frac{k_0 R_4 \Delta_3}{R_4 - R_3}\right)^{|n|}, \quad b_n^{(3)} \sim \frac{2^{1-|n|} i^{|n|} \Psi_n^{(1)}(k_0 R_2)}{(|n|-1)! k_0 R_3 \Omega_n} \left(\frac{k_0 R_4 \Delta_3}{R_4 - R_3}\right)^{|n|}, \tag{11}$$

$$a_n^{(5)} = -i b_n^{(4,5)} \sim \pi i^{|n|+1} 2^{-2|n|} \left[\frac{J_{|n|+1}(k_0 R_3) \Psi_n^{(2)}(k_0 R_2) + Y_{|n|+1}(k_0 R_3) \Psi_n^{(1)}(k_0 R_2)}{|n|!(|n|-1)! \Omega_n}\right] \\ \times \left(\frac{k_0 R_4 \Delta_3}{R_4 - R_3}\right)^{2|n|}, \tag{12}$$

where $\Lambda = J_1(k_0 R_2) Y_1(k_0 R_3) - J_1(k_0 R_3) Y_1(k_0 R_2)$ and

$$\Omega_n = \Psi_n^{(1)}(k_0 R_2) Y_{|n|-1}(k_0 R_3) + \Psi_n^{(2)}(k_0 R_2) J_{|n|-1}(k_0 R_3), \tag{13}$$

$$\Psi_n^{(1)}(x) = |n|(1-\varepsilon_1) J_{|n|}(x) + \varepsilon_1 x J_{|n|+1}(x), \quad \Psi_n^{(2)}(x) = |n|(\varepsilon_1 - 1) Y_{|n|}(x) - \varepsilon_1 x Y_{|n|+1}(x). \tag{14}$$

Note that *all* the zeroth-order coefficients in (8) and (9) do not depend on $\Delta_2$, and, as already observed for the standard cloak case (see, e.g., [10,12]), they tend *logarithmically* to zero in the limit of an ideal cloak $(\Delta_3 \to 0)$. Also the $(n \neq 0)$ th-order coefficients $a_n^{(3)}$, $b_n^{(3)}$, $b_n^{(4)}$, $a_n^{(5)}$, and $b_n^{(5)}$ in (11) and (12) do not depend on $\Delta_2$, and they tend *exponentially* to zero in the limit $\Delta_3 \to 0$. Conversely, the remaining $(n \neq 0)$ th-order coefficients $a_n^{(1)}$ and $a_n^{(2)}$ in (10) depend (exponentially) on the *ratio* $\Delta_3/\Delta_2$. This suggests that letting $\Delta_2, \Delta_3 \to 0$, while keeping their ratio *finite*, it is possible to combine the cloak and anti-cloak functions, suppressing (via the cloak) the field in the vacuum layer $R_2 < r < R_3$ (as well as the one scattered in the vacuum region $r > R_4$) and having it "restored" (via the anti-cloak) in the cylinder $r < R_1$. This justifies the previous assumption that a field "tunneling" is responsible for transmitting the field from the cloak inner interface to the anti-cloak outer interface, consistent with (2). It is worth pointing out that the field restored in the cylinder $r < R_1$ is a distorted version of the incident field (4), since the expansion coefficients (cf. (8) and (10)) are clearly different; in particular, the zeroth-order terms are vanishingly small.

## 4. Representative results

We start considering a configuration with a DPS (vacuum) inner cylinder $\left(\varepsilon_1 = \varepsilon_0, \mu_1 = \mu_0\right)$ and, consequently (in view of (2) and (3)), a DNG anti-cloak, with $R_1 = 0.4\lambda_0$, $R_2 = 0.75\lambda_0$, $R_3 = 1.7\lambda_0$, and $R_4 = 2.5\lambda_0$. We assume very slight losses (tan$\delta$=10$^{-4}$) and, for the small parameters, we choose $\Delta_2 = R_2/200$ and $\Delta_3 = R_3/200$, which yields constitutive parameters of the transformation media ranging (in absolute values) between nearly zero and 200.

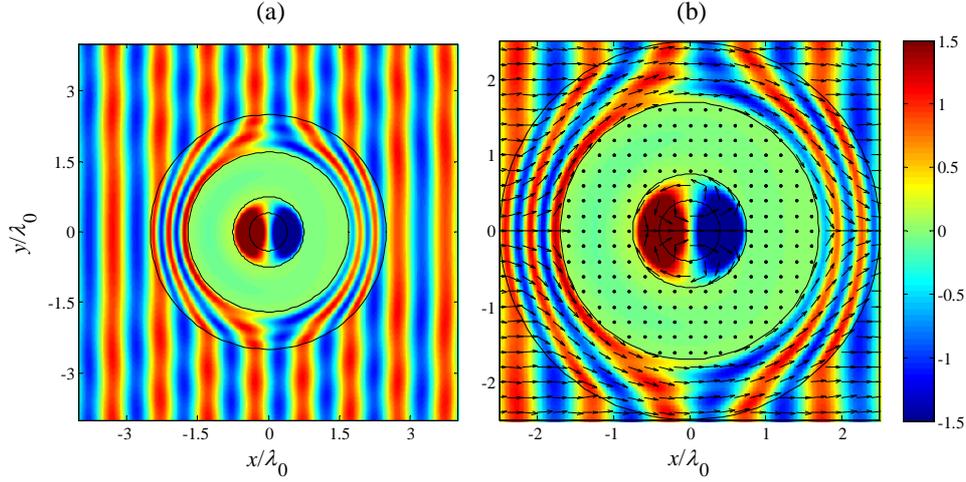

Fig. 2. (a) Magnetic field (real part) map for a configuration featuring a vacuum $\left(\varepsilon_1 = \varepsilon_0, \mu_1 = \mu_0\right)$ inner cylinder and a DNG anti-cloak, with $R_1 = 0.4\lambda_0$, $R_2 = 0.75\lambda_0$, $R_3 = 1.7\lambda_0$, $R_4 = 2.5\lambda_0$, $\Delta_2 = R_2/200$, $\Delta_3 = R_3/200$, and tan$\delta$=10$^{-4}$. (b) Magnified view with a superimposed map of the real part of the Poynting vector (normalized in the uncloaked regions, so that it is only indicative of the power flow direction).

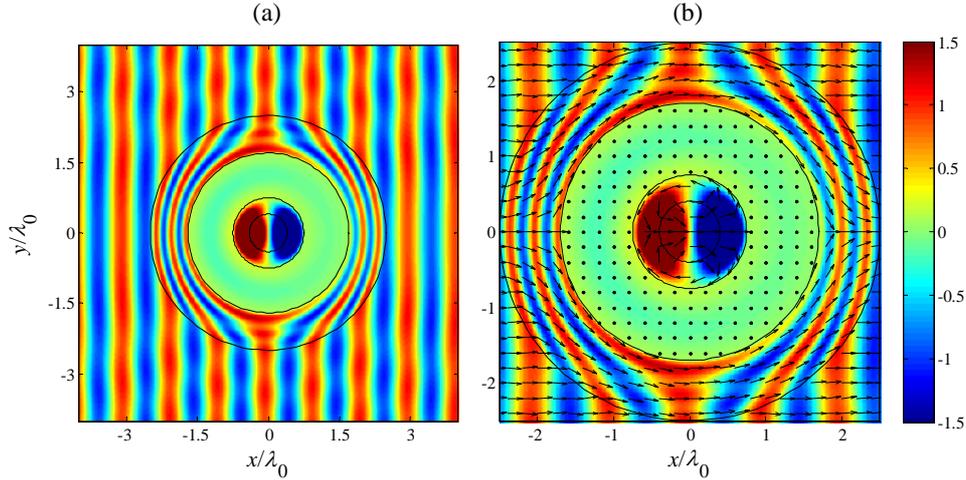

Fig. 3. As in Fig. 2, but with DNG $\left(\varepsilon_1 = -\varepsilon_0, \mu_1 = -\mu_0\right)$ inner cylinder and DPS anti-cloak.

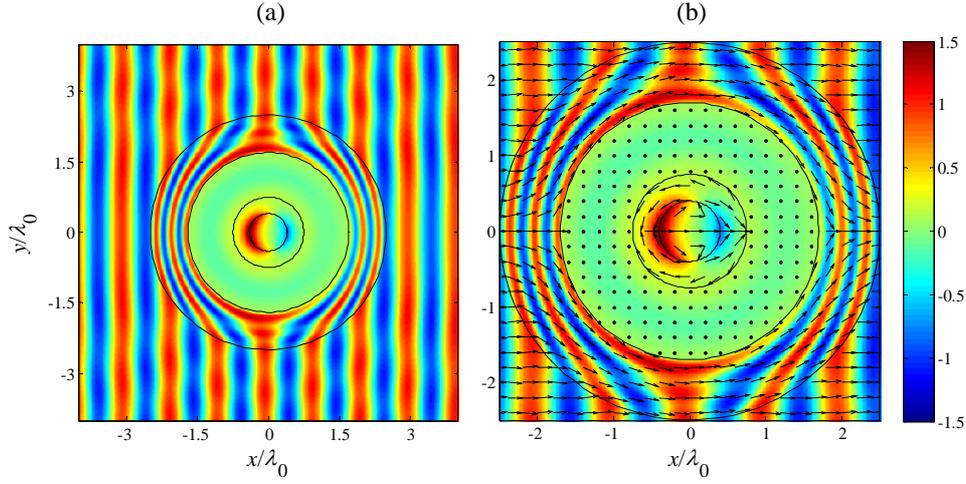

Fig. 4. As in Fig.2 , but with ENG $\left(\varepsilon_1 = -\varepsilon_0, \mu_1 = \mu_0\right)$ inner cylinder and MNG anti-cloak.

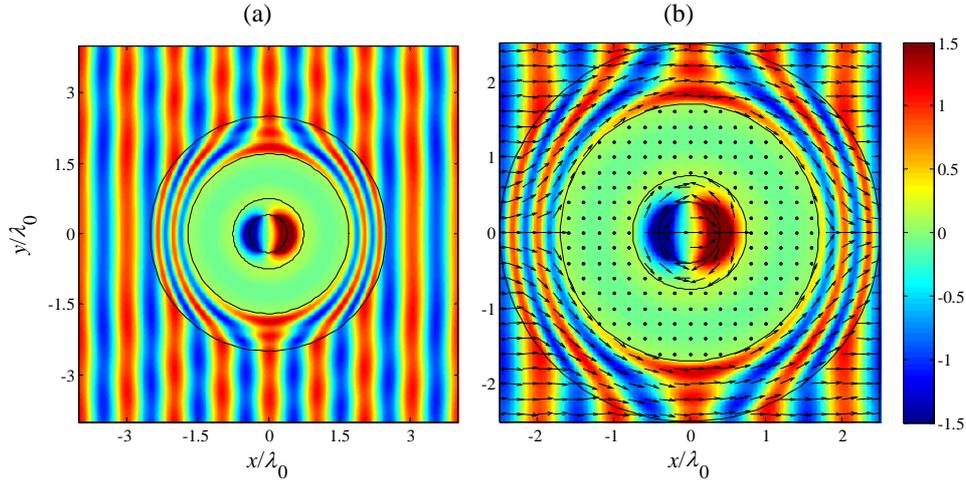

Fig. 5. As in Fig. 2, but with MNG $\left(\varepsilon_1 = \varepsilon_0, \mu_1 = -\mu_0\right)$ inner cylinder and ENG anti-cloak.

Figure 2(a) shows the real part of the magnetic field map computed via the Fourier-Bessel expansion (5), which confirms the previous observations. An external observer, placed outside the cloak ($r > R_4$), experiences a response essentially equivalent to that of a standard transformation-based cloak, with very low scattering and very mild distortion of the original planar wavefronts. On the other hand, an internal observer, placed inside the region $r < R_1$, would experience a substantially different situation, as compared with a standard cloak. In fact, the field is very weak, almost zero, in the layer $R_2 < r < R_3$, which can thus be thought as effectively cloaked, but it is restored (via the anti-cloak) in the inner cylinder $r < R_1$. The power flow may be better understood from the magnified detail (with the normalized Poynting vector map superimposed) in Fig. 2(b). Once again, outside the cloak, the picture is essentially equivalent to the standard cloak. Inside the cloak, the anti-cloak and the inner cylinder form a "resonating cavity" which, via the vanishingly small coupling through the cloaked layer is able to restore a modal field. In particular, it is interesting to observe the power circulation in the "cavity," with a forward flow in the inner cylinder, directed parallel to the impinging plane wave, which circulates back in the anti-cloak region.

Similar results may be observed for the different configurations illustrated in Figs. 3-5. Specifically, Fig. 3 shows the results for a configuration featuring a DNG $(\varepsilon_1 = -\varepsilon_0, \mu_1 = -\mu_0)$ inner cylinder and a DPS anti-cloak, for which the same qualitative considerations about the power flow, mentioned above, hold. Figure 4 corresponds to the case of an ENG $(\varepsilon_1 = -\varepsilon_0, \mu_1 = \mu_0)$ inner cylinder and an MNG anti-cloak, while Fig. 5 pertains to the dual configuration featuring an MNG $(\varepsilon_1 = \varepsilon_0, \mu_1 = -\mu_0)$ inner cylinder and an ENG anti-cloak. In these last two scenarios, the power flow inside the "cavity" exhibits a less clean-cut behavior, with the presence of "loops" at the boundary between the cylinder and the anti-cloak, possibly due to localized resonances.

The above results, which have been validated via a finite-element commercial software [20], confirm the possibility to "tunnel" the impinging wave from the cloak to a DPS, DNG or SNG anti-cloak, through the cloaked vacuum layer. Besides the interest in extending and generalizing the results in [17], the mechanisms that we have reported here are potentially very attractive from the application viewpoint, for essentially two reasons:

i) They may relax some of the practical-feasibility limitations of the DNG anti-cloak introduced in [17]. While it is a formidable technological challenge to synthesize DNG metamaterials with the high (absolute) values of the constitutive parameters required by (3), it is certainly easier to deal with the DPS or SNG alternative configurations introduced here. In particular, the DPS anti-cloak configuration entails the same technological complications as a standard cloak; in this case, the DNG character of the inner cylinder poses a less demanding technological challenge, since its constitutive parameters $(\varepsilon_1 = -\varepsilon_0, \mu_1 = -\mu_0)$ are uniform and finite.

ii) They suggest interesting scenarios for applications, where one may be able to cloak a region of space (the vacuum shell) and yet maintain the capability of "sensing" the outside field from the inside by creating, via the anti-cloak, an "invisible observation window."

Although the present study has not been focused on practical applications, it is nevertheless interesting to explore the effects of the presence of objects in the cloaked layer, and of the main nonidealities, namely, the (small but finite) parameters $\Delta_2$ and $\Delta_3$, as well as the unavoidable presence of material losses. In this sense, we report here a parametric study in terms of the two non-dimensional parameters:

$$Q_e = \frac{2}{\pi} \sum_{m=-\infty}^{\infty} |a_n^{(5)}|^2, \quad Q_i = \left(\frac{R_1^2}{R_3^2 - R_2^2}\right) \frac{\int_{R_2}^{R_3} \int_0^{2\pi} |H_z(r,\phi)|^2 rdrd\phi}{\int_0^{R_1} \int_0^{2\pi} |H_z(r,\phi)|^2 rdrd\phi}. \qquad (15)$$

The *exterior* parameter $Q_e$ is readily recognized to be the total scattering cross-sectional width per unit length (normalized by the vacuum wavelength), which quantifies the visibility of the overall configuration to a far-field exterior observer, whereas the *interior* parameter $Q_i$ quantifies the capability of coupling the field inside the inner cylinder $r < R_1$ while maintaining a very weak intensity in the (ideally cloaked) layer $R_2 < r < R_3$. In the ideal case ($\Delta_2, \Delta_3 \to 0$ and lossless materials), both parameters should vanish.

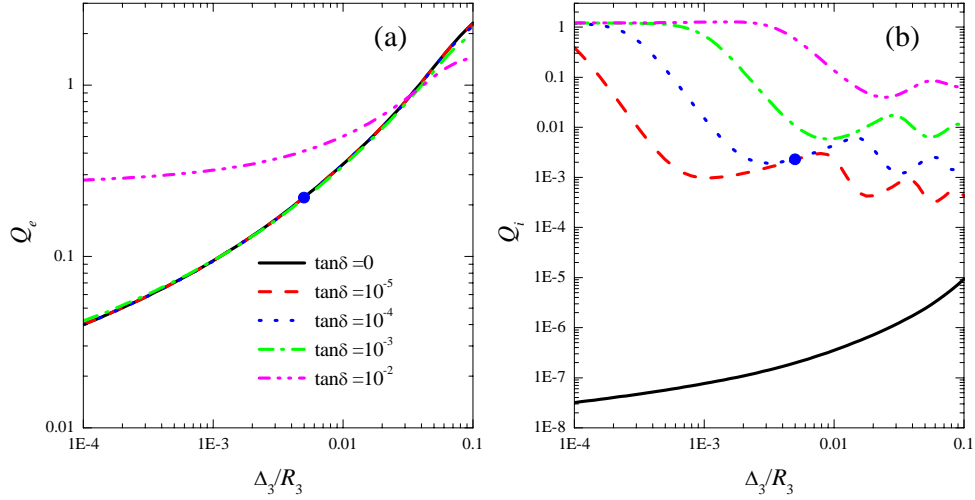

Fig. 6. Geometry as in Fig. 2 (DPS inner cylinder, DNG anti-cloak). Exterior (a) and interior (b) parameters in (15) as a function of the parameter $\Delta_3/R_3$, for a fixed ratio $\Delta_2/\Delta_3 = R_2/R_3 \approx 0.44$ and different values of the loss-tangent. The blue dots indicate the values corresponding to the geometry of Fig. 2.

Figure 6(a) shows, for the DNG anti-cloak configuration of Fig. 2, the behavior of the two parameters in (15) as a function of the small parameter $\Delta_3$ (for a fixed ratio $\Delta_2/\Delta_3 = R_2/R_3 \approx 0.44$), and for various values of the loss-tangent ranging from zero to $10^{-2}$. In particular, the blue dots denote the level of losses corresponding to the example in Fig. 2. For the lossless case, one observes the anticipated monotonic reduction of both parameters as $\Delta_3 \to 0$. The scattering width $Q_e$ (see Fig. 6(a)) turns out to be only mildly dependent on the losses (with a significant departure from the lossless behavior observable for $\tan\delta \sim 10^{-2}$), and, as also observed in the standard cloak case, essentially decreases monotonically with the parameter $\Delta_3$. The effect of losses is much more evident in the interior parameter $Q_i$ (see Fig. 6(b)). In particular, increasing the losses, one observes the appearance (for small values of $\Delta_3$) and progressive enlargement of regions where $Q_i$ can be as high as ~1, thereby implying that the average values of the field intensity in the inner cylinder and in the cloaked layer are actually comparable. In such regions, while the exterior visibility of the configuration can still be relatively small (like in the standard cloak case), the peculiar coupling (via "tunneling") effects observed in Figs. 2-5 are effectively destroyed. However, outside these ranges of parameters, $Q_i$ decreases with an *oscillating* behavior, exhibiting minima that may be acceptably small (~$10^{-2}$) even in the presence of moderate losses ($\tan\delta \sim 10^{-2}$). This leads to the conclusion that, in the presence of losses, the "ideal" condition $\Delta_3 \to 0$ is not necessarily "optimal," and that a suitable tradeoff between the two parameters can be achieved for *finite* values of $\Delta_3$. This is expected, since the ideal configuration $\Delta_2, \Delta_3 \to 0$ would create a barrier for the electromagnetic fields at the inner boundary of the cloak, that necessitates to be relaxed when losses are present. The resonant nature of this cloak/anti-cloak interaction and of the field tunneling described above is clearly evident, which indeed may be significantly affected by losses and parameter deviations. Qualitatively similar trends, not shown here for sake of brevity, are observed for the other cloak/anti-cloak combinations (cf. Figs. 3-5).

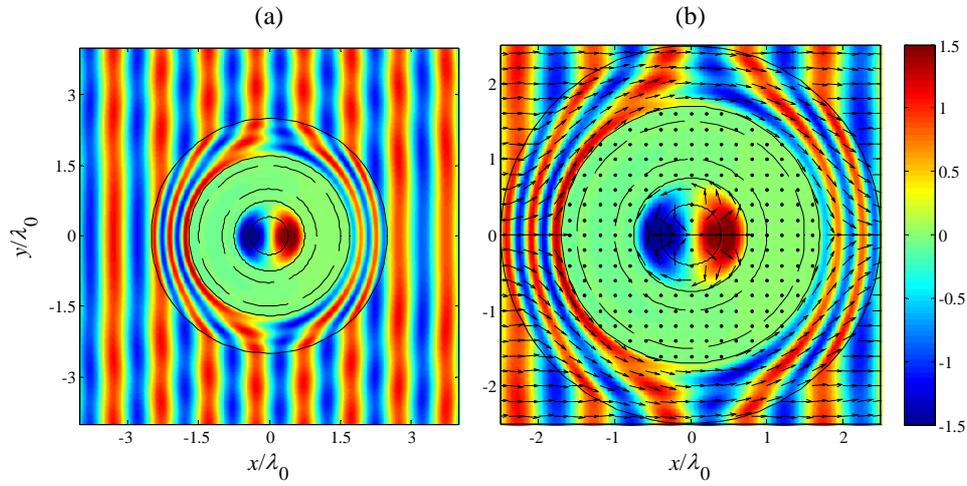

Fig. 7. As in Fig. 2, but with a dielectric coaxial annular layer of radii $R_a = \lambda_0$ and $R_b = 1.5\lambda_0$ (shown dashed) and permittivity $\varepsilon_{obj} = 2\varepsilon_0$ $\left(\tan\delta = 10^{-4}\right)$ inside the cloaked layer.

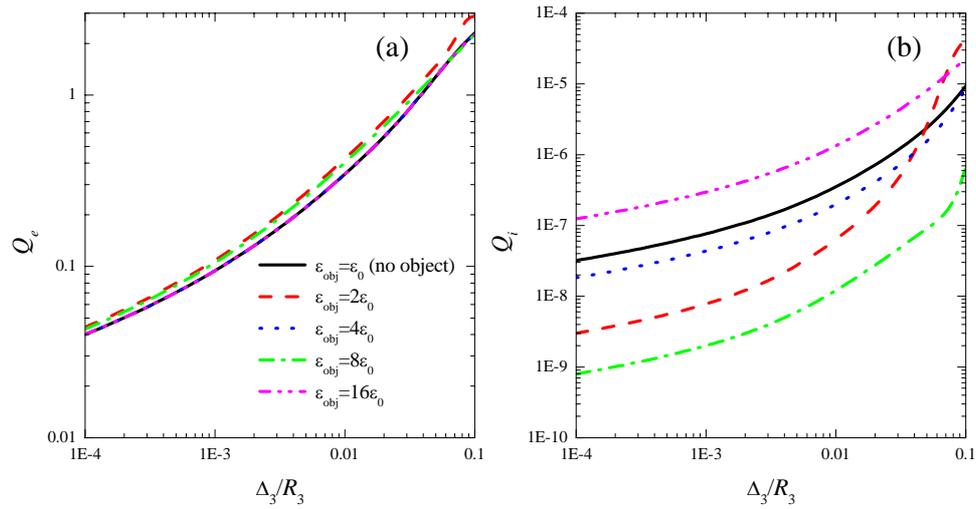

Fig. 8. As in Fig. 6, but in the absence of losses, and in the presence of a dielectric coaxial annular layer of radii $R_a = \lambda_0$ and $R_b = 1.5\lambda_0$ inside the cloaked layer, for various values of the object permittivity $\varepsilon_{obj}$.

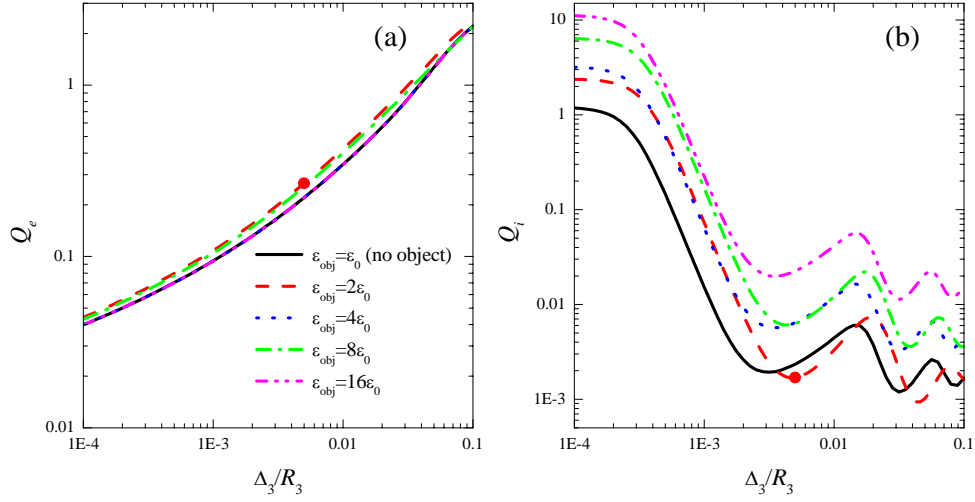

Fig. 9. As in Fig. 8, but for $\tan\delta = 10^{-4}$. The red dots indicate the values corresponding to the geometry of Fig. 7.

We also considered the possible presence of penetrable objects inside the cloaked layer. Referring to the scenario and parameters of Fig. 2, Fig. 7 shows the field maps in the presence of a dielectric coaxial annular layer of radii $R_a = \lambda_0$ and $R_b = 1.5\lambda_0$ (shown dashed), and permittivity $\varepsilon_{obj} = 2\varepsilon_0$ ($\tan\delta = 10^{-4}$). The coaxial annular shell geometry was chosen in order to preserve the analytical tractability of the problem (via straightforward generalization of (5)). As one can see, also by comparison with Fig. 2, the dielectric annular layer is effectively cloaked. From a topological viewpoint, this may also be considered as the first example of *selective* cloaking of a *multiply-connected* object.

For more quantitative assessments, we studied the behavior of the interior and exterior parameters in (15) for different values of the object permittivity and levels of losses. Figure 8 shows the results in the ideal lossless case, which confirm the expected monotonic decreasing of both parameters as $\Delta_3 \to 0$. In particular, the scattering width $Q_e$ depends only mildly on the object parameters, whereas the interior parameter $Q_i$, while remaining always rather small ($<10^{-4}$), exhibits a stronger dependence. Finally, Fig. 9 shows the more realistic results pertaining to a slightly-lossy case ($\tan\delta = 10^{-4}$, as in Fig. 7). As for the case in the absence of the object (cf. Fig. 6), the scattering width $Q_e$ is not sensibly affected, whereas the effects are more dramatic for the interior parameter $Q_i$, which exhibit a decreasing oscillatory behavior, with local minima which may still be acceptably small. These minima tend to increase with increasing the object permittivity, as one could expect by observing that the cloak/anti-cloak interaction should vanish in the limit of an impenetrable layer.

## 5. Conclusions and outlook

In this paper, elaborating upon the recently introduced concept of "anti-cloak" [17], we have explored the peculiar coupling effects that may be obtained by pairing a cloak and an anti-cloak separated by a vacuum layer, surrounding a dielectric or metamaterial cylinder. We have shown that, depending on the constitutive parameters of the inner cylinder, four distinct (including DPS and SNG) configurations for an anti-cloak are possible, which can relax some of the practical-feasibility limitations of the DNG anti-cloak introduced in [17]. Besides the potential application (as a cloak "countermeasure") proposed in [17], we have suggested interesting application scenarios for which a region of space may be cloaked, while maintaining the capability of somehow "sensing" the outside field from the inside. Moreover, via a full-wave analytical study, we have explored the effects of the presence of penetrable objects inside the cloaked region, as well as the unavoidable non-idealities in the constitutive parameters and presence of losses.

The results obtained in this study may pave the way to new exciting developments in cloaking applications. In this sense, current and future studies are aimed at the parametric optimization of the proposed configuration, as well as the exploration of other possible interactions of interest.